\documentclass{article}
\usepackage{spconf,amsmath,graphicx}

\usepackage{enumitem}
\setlist{nosep, leftmargin=14pt}

\usepackage{mwe} 
\usepackage{xcolor}
\usepackage{multirow}
\usepackage{amssymb}

\title{Focal Attention Networks: Optimising Attention for Biomedical Image Segmentation}
\makeatletter
\def\@name{ \emph{Michael Yeung}$^{1,2,3}$, \emph{Leonardo Rundo}$^{2,4}$, \emph{Evis Sala}$^{2,4}$, \emph{Carola-Bibiane Sch\"{o}nlieb}$^{5}$, \emph{Guang Yang}$^{3}$\\}
\makeatother

\address{$^1$ School of Clinical Medicine, University of Cambridge, Cambridge, UK\\
$^2$ Department of Radiology, University of Cambridge, Cambridge, UK\\
$^3$ National Heart \& Lung Institute, Imperial College London, London, UK \\
$^4$ Cancer Research UK Cambridge Centre, University of Cambridge, Cambridge, UK\\
$^5$ Department of Applied Mathematics and Theoretical Physics, University of Cambridge, Cambridge, UK\\
}
\begin{document}
%
\maketitle
\begin{abstract}
In recent years, there has been increasing interest to incorporate attention into deep learning architectures for biomedical image segmentation. The modular design of attention mechanisms enables flexible integration into convolutional neural network architectures, such as the U-Net. Whether attention is appropriate to use, what type of attention to use, and where in the network to incorporate attention modules, are all important considerations that are currently overlooked. In this paper, we investigate the role of the Focal parameter in modulating attention, revealing a link between attention in loss functions and networks. By incorporating a \textit{Focal distance penalty term}, we extend the Unified Focal loss framework to include boundary-based losses. Furthermore, we develop a simple and interpretable, dataset and model-specific heuristic to integrate the Focal parameter into the Squeeze-and-Excitation block and Attention Gate, achieving optimal performance with fewer number of attention modules on three well-validated biomedical imaging datasets, suggesting judicious use of attention modules results in better performance and efficiency.
\end{abstract}
\section{Introduction}
\label{sec:intro}

Attention provides neural networks with the capacity to selectively process salient inputs. In the context of image segmentation, attention mechanisms may be broadly divided into two complementary types: spatial attention, which enhances processing of salient locations within the image, and channel attention, which calibrates feature maps based on relative importance \cite{schlemper2019attention, hu2018squeeze,  yeung2021focus, roy2018concurrent, woo2018cbam}. Both forms of attention may be encapsulated into modules, which enable flexible integration into existing convolutional neural network (CNN) architectures. The most widely used CNN for biomedical image segmentation is the U-Net, consisting of a symmetrical encoder-decoder structure with skip connections \cite{ronneberger2015u}. Attention-based variants of the U-Net include the Attention U-Net, which incorporates spatial attention using the Attention Gate (AG), and the USE-Net, which uses the channel-based Squeeze-and-Excitation (SE) block \cite{schlemper2019attention, hu2018squeeze, rundo2019use}.  

Despite the widespread use of attention in CNNs for image segmentation, it remains unclear when attention is appropriate to use, which type of attention to use, and where the optimal location is for use within the network. Currently, performance benchmarking is the only method available for evaluating the value of attention. However, the contribution of individual attention modules cannot be inferred, and even with ablation studies, only a minor subset of all positional combinations can be reasonably evaluated. Without understanding how individual attention modules affect performance, it is not possible to determine where to optimally place attention modules.

The main contributions of this work may be summarised as follows:
\begin{enumerate}
\item We leverage the Focal parameter to modulate attention across both loss function and network contexts, incorporating a Focal Distance Penalty Term to further generalise the Unified Focal loss framework, and integrating a Focal layer into network attention modules.
\item We demonstrate consistently improved performance using the Unified Focal loss and Focal Attention networks across three, well-validated open-source biomedical imaging datasets.
\item We develop simple and interpretable, model and dataset-specific heuristics for deciding the optimal type, location and strength of attention, and which will facilitate further, large-scale benchmarking studies.
\end{enumerate}

\section{Materials and methods}

\subsection{Focal Attention in loss functions}

The Focal loss was designed as a variant of the cross-entropy loss to address class imbalanced datasets for classification, by selectively downweighting well-classified examples \cite{Lin_2017_ICCV}. While classification is interested in class imbalance at the image level, image segmentation deals with class imbalance at the pixel level, and in biomedical image segmentation, class imbalance is frequently observed with objects such as cell nuclei or tumours which occupy a small area relative to the image. The Unified Focal loss framework generalises distribution-based and region-based loss functions to handle class imbalanced datasets \cite{yeung2021unified}. For a given softmax output for classes $c$, where $r$ and $t$ refer to the rare class and ground truth respectively, the Unified Focal loss ($\mathcal{L}_\text{UF}$) is defined as the weighted sum of the Asymmetric Focal loss ($\mathcal{L}_\text{AF}$) and Asymmetric Focal Tversky loss ($\mathcal{L}_\text{AFT}$):

\begin{equation}
\mathcal{L}_\text{UF}=\lambda \mathcal{L}_\text{AF}+(1-\lambda) \mathcal{L}_\text{AFT},
\label{eq:GFL}
\end{equation}
where:

\begin{equation}
\mathcal{L}_\text{AF}=- \frac{\delta}{N} y_{i: r} \log \left(p_{t, r}\right)-\frac{1-\delta}{N} \sum_{c \neq r}\left(1-p_{t, c}\right)^{\gamma} \log \left(p_{t, r}\right),
\end{equation}

\begin{equation}
\mathcal{L}_{\mathrm{AFT}}=\sum_{\mathrm{c} \neq \mathrm{r}}(1-\mathrm{TI})+\sum_{\mathrm{c}=\mathrm{r}}(1-\mathrm{TI})^{1-\gamma}.
\end{equation}

The Tversky Index (TI) is defined as:

\begin{equation}
\operatorname{TI} =
\frac{\sum_{i=1}^{N} p_{0 i} g_{0 i}}{\sum_{i=1}^{N} p_{0 i} g_{0 i}+\delta\sum_{i=1}^{N} p_{0 i} g_{1 i}+(1-\delta) \sum_{i=1}^{N} p_{1 i} g_{0 i}},
\label{eq:modified_Tversky_index}
\end{equation}
where $p_{0 i}$ is the probability of pixel $i$ belonging to the foreground class i.e. segmentation target and $p_{1 i}$ is the probability of pixel belonging to background class. $g_{0 i}$ is $1$ for foreground and 0 for background and conversely $g\textsubscript{1i}$ takes values of $1$ for background and 0 for foreground. 

The three hyperparameters in the Unified Focal loss are $\lambda$, which controls the relative weights of the two component losses, $\delta$, which controls the relative weighting of positive and negative examples, and $\gamma$, which controls the relative weighting of easy and difficult examples. 

Another class of loss functions are boundary-based loss functions, which compute Distance Transform Maps (DTM) based on Euclidean distances to penalise predictions relative to class boundaries \cite{kervadec2019boundary}. The Distance Penalty Term (DPT) is defined as the inverse of DTM, penalising incorrect predictions close to boundaries \cite{caliva2019distance, sugino2021loss}. Here, we generalise the DPT, defining the Focal Distance Penalty Term (FDPT) by applying a Focal parameter, $\epsilon$:

\begin{equation}
\mathcal{W}_{c}^{\mathrm{FDPT}} =  (\mathcal{W}_{c}^{\mathrm{DPT}})^{\epsilon}.
\end{equation}

\begin{figure}[ht!]
    \includegraphics[width=0.5\textwidth]{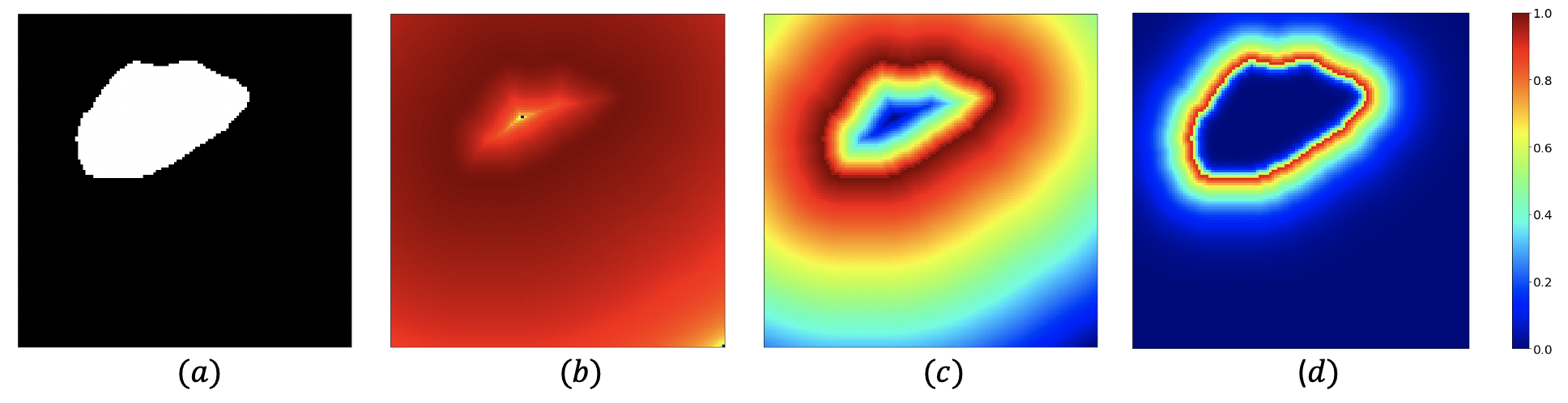}
    \caption{Focal Distance Penalty Term distance maps visualised as heatmaps where (a) label image, (b) $\epsilon = 0.1$, (c) $\epsilon = 1$ (equivalent to DPT) and (d) $\epsilon = 10$.}
    \label{fig:figure_1}
\end{figure}

The FDPT establishes a continuity between no boundary awareness ($\epsilon = 0$), to varying degrees of boundary awareness, revealing that the DPT is a particular solution where $\epsilon = 1$ (Fig.~\ref{fig:figure_1}).

\begin{figure}[ht!]
    \centering
    \includegraphics[width=0.45\textwidth]{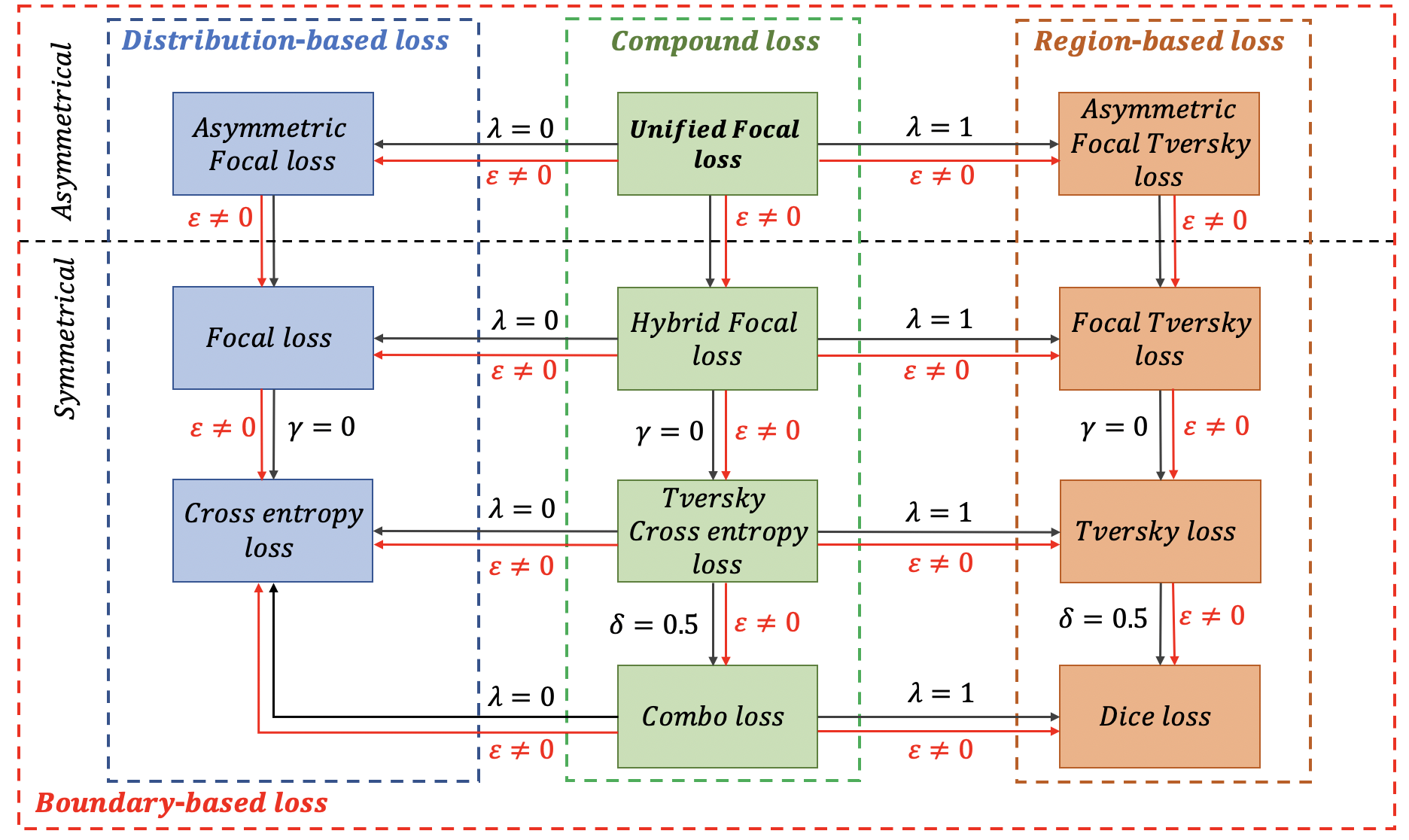}
    \caption{The extended Unified Focal loss framework generalises numerous distribution-based, region-based and boundary-based loss functions. By fixing  hyperparameter values, particular loss functions may be derived as specified by the arrows.}
    \label{fig:figure_2}
\end{figure}

We integrate the FDPT into the Unified Focal loss framework by replacing the ground truth with its respective FDPT, resulting in a loss function where optimisation involves selective attention to difficult to segment regions and boundaries (Fig.~\ref{fig:figure_2}). An important consequence of generalising loss functions is that the performance of the extended Unified Focal loss is, with the appropriate hyperparameter setting, guaranteed to perform at least equivalent, if not better, than any of its component loss functions.

\subsection{Focal Attention in networks}

\begin{figure}[ht!]
    \centering
    \includegraphics[width=0.37\textwidth]{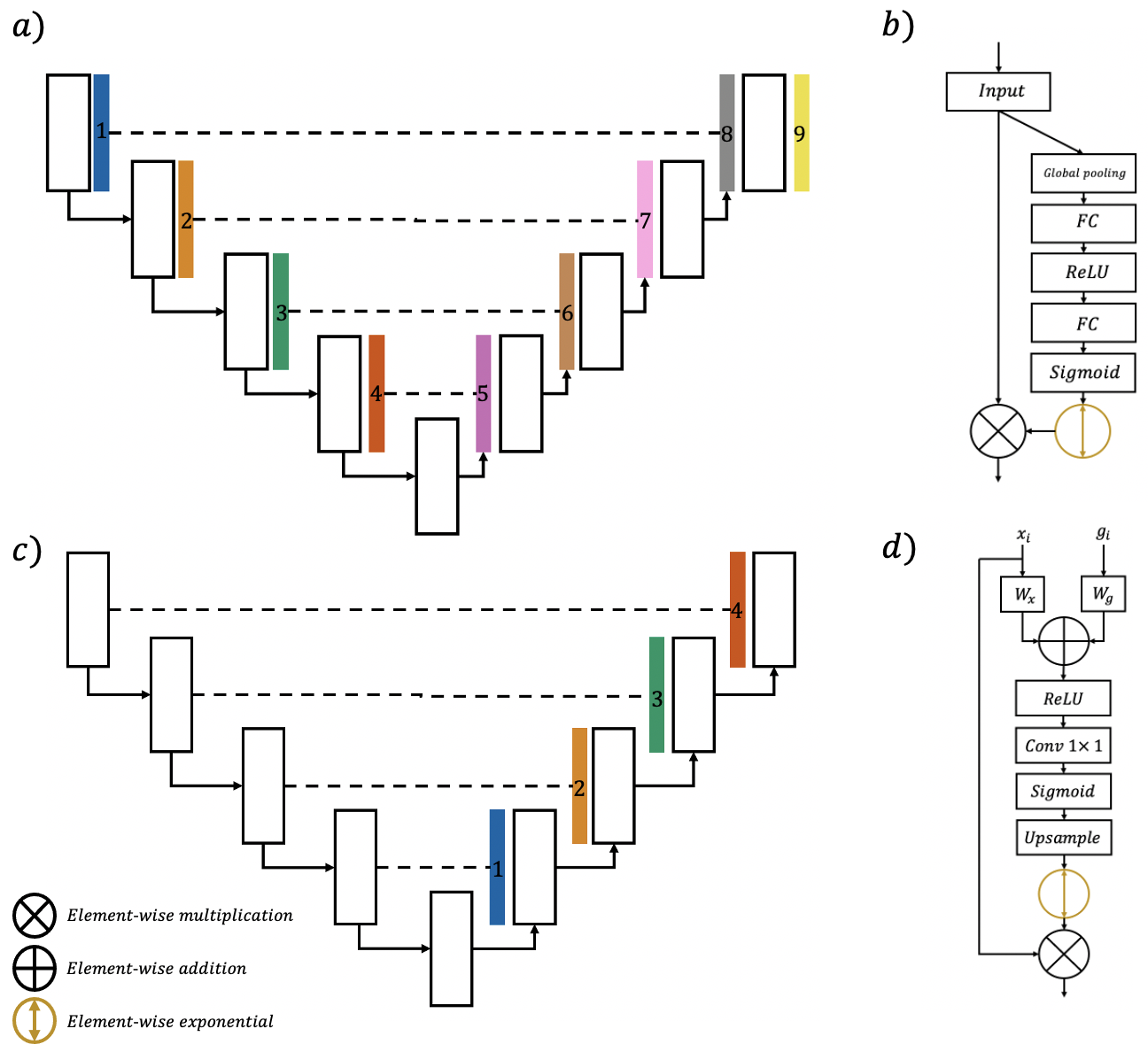}
    \caption{Simplified diagrams of the (a) USE-Net, (b) SE block, (c) Attention U-Net and (d) AG. The coloured blocks and numbering indicate the position of attention modules.}
    \label{fig:figure_3}
\end{figure}

Similar to loss functions, the Focal parameter generalises network attention mechanisms by modulating attention strength \cite{yeung2021focus}. As examples, we select the SE block for channel attention and the AG for spatial attention, which has been integrated into the U-Net architecture in the USE-Net and Attention U-Net, respectively (Fig.~\ref{fig:figure_3}) \cite{schlemper2019attention, rundo2019use}. Briefly, SE blocks achieve channel attention by performing initial feature aggregation using global average pooling along the spatial axis (`squeeze'), followed by two fully connected (FC) layers with ReLU and sigmoid activations producing the `excitation' operation. In contrast, the AG uses contextual information from the gating signal $g_{x}$ to prune skip connection $x_{i}$. The Focal layer is inserted after the generation of attention weights and prior to recalibration, and involves a single, trainable parameter, that modulates attention strength. By initialising the Focal parameter to one, the Focal Attention module initially behaves identical to its respective attention module, but during training, the attention strength is optimised through parameter updating using backpropagation. 

We investigate a further use of the Focal layer to determine usefulness of individual attention modules. By initialising the Focal parameter to zero and monitoring the Focal layer weights during training, we expect attention modules that are either neutral or harmful to performance to remain close to zero.  There are no definitive cut-off values for the Focal parameter to distinguish likely useful from neutral or harmful attention modules, and depends on weighing potential performance benefits against efficiency costs. For this preliminary study, we select $0.2$ as the threshold below which we remove attention modules at convergence. 

\subsection{Dataset descriptions and evaluation metrics}

To evaluate the extended Unified Focal loss and Focal Attention networks, we select three well-validated, open-source biomedical imaging datasets: Digital Retinal Images for Vessel Extraction (DRIVE), 2018 Data Science Bowl (2018DSB) and CVC-ClinicDB \cite{staal2004ridge, caicedo2019nucleus, bernal2015wm}. Briefly, the DRIVE dataset consists of 40 coloured fundus photographs for retinal vessel segmentation, 2018DSB comprises 670 light microscopy images for nuclei segmentation, and the CVC-ClinicDB dataset consists of 612 frames containing polyps obtained during optical colonoscopy. A summary of the datasets and training details are presented in Table \ref{tab:table1}. 

\begin{table}[ht!]
\centering
\caption{Details of datasets and training setup used for our experiments. * The effective \%Foreground is significantly higher if only the field of view is considered.}
\scalebox{0.45}{
\begin{tabular}{lcccccccc}
\hline
Dataset      & Segmentation     & \#Images & Size      & \#Training & \#Validation & \#Test & \%Foreground & UFL $\gamma$ \\
\hline
DRIVE        & Retinal vessel   & 40       & $512 \times 512  \times 3$ & 16         & 4            & 20     & 8.70*        & 0.1     \\
2018DSB      & Cell nuclei      & 670      & $256 \times 256  \times 3$ & 428        & 108          & 134    & 14.5         & 0.2     \\
CVC-ClinicDB & Colorectal polyp & 612      & $288 \times 384  \times 3$ & 392        & 98           & 122    & 9.30         & 0.3 
\\ \hline
\end{tabular}}
\label{tab:table1}
\end{table}

For evaluation, we calculate Dice Similarity Coefficient (DSC), precision and recall metrics per image and average over the independent test set.

\subsection{Implementation details}

For our experiments, we used the Medical Image Segmentation with Convolutional Neural Networks (MIScnn) open-source Python library \cite{muller2021miscnn}. Our implementations made use of Keras with Tensorflow backend and were trained using NVIDIA P100 GPUs. For all experiments, except for the DRIVE dataset that is already partitioned into 20 training images and 20 testing images, we randomly partitioned each dataset into 80\% development and 20\% test set, and further divided the development set into 80\% training set and 20\% validation set. All images were normalised to $[0,1]$ using the z-score, and we used the following data augmentation: scaling, rotation, mirroring, elastic deformation and brightness.  

Model parameters were initialised using the Xavier initialisation.  We trained each model using the Adam optimizer with an initial learning rate of $1\times10^{-3}$ and used ReduceLROnPlateau to reduce the learning rate by 0.1 if the validation loss did not improve after 25 epochs, and the EarlyStopping callback to terminate training if the validation loss did not improve after 50 epochs. 

For the Unified Focal loss, we set $\delta = 0.6$ and $\lambda = 0.5$, and empirically determined the optimal $\gamma$ value for each dataset \cite{yeung2021unified}. The FDPT was empirically set with $\epsilon = 0.1$. For the SE block, we set the reduction ratio $r = 8$ \cite{hu2018squeeze, rundo2019use}. For all experiments, we use instance normalisation, and train with a batch size of 1 \cite{ronneberger2015u, zhou2019normalization}.

\section{Results}

\subsection{Loss function experiments}

We compare the performance of the U-Net when optimised with the DSC loss, DSC + cross entropy (CE) loss, and UFL, with the FDPT set either to $\epsilon = 0$ (equivalent to no penalty), $\epsilon = 1$ (equivalent to DPT) and $\epsilon = 0.1$. The results for the loss function comparisons on the three datasets are shown in Table \ref{tab:table2}. 

The UFL + FDPT is associated with the highest DSC values across all three datasets, with a DSC of 0.8155, 0.9165 and 0.8993 for the DRIVE, 2018DSB and CVC-ClinicDB respectively. The UFL achieves consistently higher DSC scores compared to the DSC and DSC + CE losses. Worse performance was often observed with the DPT compared to no penalty, suggesting the DPT may focus too strictly on boundaries, while better performance was observed when the strength of boundary attention was reduced by setting a lower $\epsilon$ value. 

\begin{table}[ht!]
\centering
\caption{Performance comparisons using DSC loss, DSC + CE loss and UFL loss, with and without boundary attention, on DRIVE, 2018DSB and CVC-ClinicDB. The highest scores are denoted in bold.}
\scalebox{0.35}{
\begin{tabular}{llccccccccc}
\hline
Dataset                       & \multicolumn{1}{c}{Metrics} & DSC              & DSC + DPT & DSC + FDPT & DSC + CE & DSC + CE + DPT   & DSC + CE + FDPT  & UFL             & UFL + DPT       & UFL + FDPT      \\
\hline
\multirow{3}{*}{DRIVE}        & DSC                         & 0.8082          & 0.8086   & 0.8105    & 0.8093  & 0.8070          & 0.8116          & 0.8142          & 0.8142          & \textbf{0.8155} \\
                              & Precision                   & 0.8473          & 0.8498   & 0.8440    & 0.8480  & \textbf{0.8596} & 0.8468          & 0.8199          & 0.8298          & 0.8075          \\
                              & Recall                      & 0.7766          & 0.7751   & 0.7836    & 0.7776  & 0.7640          & 0.7829          & 0.8127          & 0.8031          & \textbf{0.8276} \\
                              \hline
\multirow{3}{*}{2018DSB}      & DSC                         & 0.9147          & 0.9016   & 0.9150    & 0.9148  & 0.9085          & 0.9159          & 0.9157          & 0.9129          & \textbf{0.9165} \\
                              & Precision                   & \textbf{0.9205} & 0.9230   & 0.9191    & 0.9140  & 0.9236          & 0.9163          & 0.9061          & 0.9008          & 0.9196          \\
                              & Recall                      & 0.9168          & 0.8891   & 0.9184    & 0.9233  & 0.9012          & 0.9231          & 0.9324          & \textbf{0.9327} & 0.9204          \\
                              \hline
\multirow{3}{*}{CVC-ClinicDB} & DSC                         & 0.8826          & 0.8174   & 0.8833    & 0.8917  & 0.8536          & \textbf{0.8993} & 0.8937          & 0.8622          & \textbf{0.8993} \\
                              & Precision                   & \textbf{0.9175} & 0.9058   & 0.9117    & 0.9166  & 0.9135          & 0.9155          & 0.8965          & 0.8913          & 0.9074          \\
                              & Recall                      & 0.8759          & 0.7614   & 0.8816    & 0.8874  & 0.8171          & 0.9024          & \textbf{0.9096} & 0.8563          & 0.9092       
                              \\ \hline
\end{tabular}}
\label{tab:table2}
\end{table}

\subsection{Focal Attention network experiments}

The results using Focal SE blocks are shown in Table \ref{tab:table3}. Integrating Focal variants of the SE block achieved the highest DSC score. Interestingly, the highest DSC score for the DRIVE and 2018DSB dataset was observed after attention module selection, suggesting certain attention modules that were removed likely worsened performance.

\begin{table}[ht!]
\centering
\caption{Performance comparisons using the U-Net, USE-Net and its Focal variant, before and after attention module selection. The highest scores are denoted in bold.}
\scalebox{0.45}{
\begin{tabular}{lccccc}
\hline
 &           & U-Net           & USE-Net & Focal USE-Net   & Focal USE-Net   \\
 \hline
                                                & \#SE      & 0               & 9       & 9               & 2               \\
                                                \hline
                                                & DSC       & 0.8142          & 0.8145  & 0.8152          & \textbf{0.8159} \\
                                                & Precision & 0.8199          & 0.8252  & 0.8056          & 0.8114          \\
\multirow{-4}{*}{DRIVE}                         & Recall    & 0.8127          & 0.8079  & \textbf{0.8289} & 0.8246          \\
\hline
                                                & \#SE      & 0               & 9       & 9               & 4               \\
                                                \hline
                                                & DSC       & 0.9157          & 0.9131  & 0.9159          & \textbf{0.9179} \\
                                                & Precision & 0.9061          & 0.9001  & 0.9024          & \textbf{0.9085} \\
\multirow{-4}{*}{2018 DSB}                      & Recall    & 0.9324          & 0.9354  & \textbf{0.9374} & 0.9343          \\
\hline
                                                & \#SE      & 0               & 9       & 9               & 5               \\
                                                \hline
                                                & DSC       & 0.8937          & 0.8952  & \textbf{0.9005} & 0.8952          \\
                                                & Precision & 0.8965          & 0.9084  & 0.9097          & \textbf{0.9111} \\
\multirow{-4}{*}{CVC-ClinicDB}                  & Recall    & \textbf{0.9096} & 0.9023  & 0.9032          & 0.9059 
\\ \hline
\end{tabular}}
\label{tab:table3}
\end{table}

The results using Focal AG are shown in Table \ref{tab:table4}. The performance improvements were less consistent than with the USE-Net. For the DRIVE dataset, the best performance was observed with the U-Net, although better performance was observed with the Focal AG in the 2018DSB and CVC-ClinicDB dataset. This may be expected, given that the retinal vessels in the DRIVE dataset extend to cover the entire field of view, making spatial attention effectively redundant for this task.

\begin{table}[ht!]
\centering
\caption{Performance comparisons using the U-Net, Attention U-Net and its Focal variant, before and after attention module selection. The highest scores are denoted in bold.}
\scalebox{0.45}{
\begin{tabular}{lccccc}
\hline
  &           & U-Net           & \begin{tabular}[c]{@{}c@{}}Attention\\ U-Net\end{tabular} & \begin{tabular}[c]{@{}c@{}}Focal Attention\\ U-Net\end{tabular} & \begin{tabular}[c]{@{}c@{}}Focal Attention\\ U-Net\end{tabular} \\
\hline
                                                & \#AG      & 0               & 4                                                         & 4                                                               & 0                                                               \\ \hline
                                                & DSC       & \textbf{0.8142} & 0.8138                                                    & 0.8138                                                          & \textbf{-}                                                      \\
                                                & Precision & \textbf{0.8199} & 0.8049                                                    & 0.8005                                                          & -                                                               \\
                                                
\multirow{-4}{*}{DRIVE}                         & Recall    & 0.8127          & 0.8274                                                    & \textbf{0.8325}                                                 & -                                                               \\\hline
                                                & \#AG      & 0               & 4                                                         & 4                                                               & 1                                                               \\ \hline
                                                & DSC       & 0.9157          & 0.9127                                                    & 0.9145                                                          & \textbf{0.9158}                                                 \\
                                                & Precision & 0.9061          & 0.8997                                                    & 0.9018                                                          & \textbf{0.9057}                                                 \\ 
                                                
\multirow{-4}{*}{2018 DSB}                      & Recall    & 0.9324          & 0.9345                                                    & \textbf{0.9365}                                                 & 0.9341                                                          \\ \hline
                                                & \#AG      & 0               & 4                                                         & 4                                                               & 2                                                               \\ \hline
                                                & DSC       & 0.8937          & 0.9051                                                    & 0.9063                                                          & \textbf{0.9118}                                                 \\
                                                & Precision & 0.8965          & 0.9185                                                    & 0.9134                                                          & \textbf{0.9195}                                                 \\
\multirow{-4}{*}{CVC-ClinicDB}                  & Recall    & 0.9096          & 0.9078                                                    & 0.9126                                                          & \textbf{0.9220}                         \\ \hline                       
\end{tabular}}
\label{tab:table4}
\end{table}

\begin{figure}[ht!]
    \centering
    \includegraphics[width=0.41\textwidth]{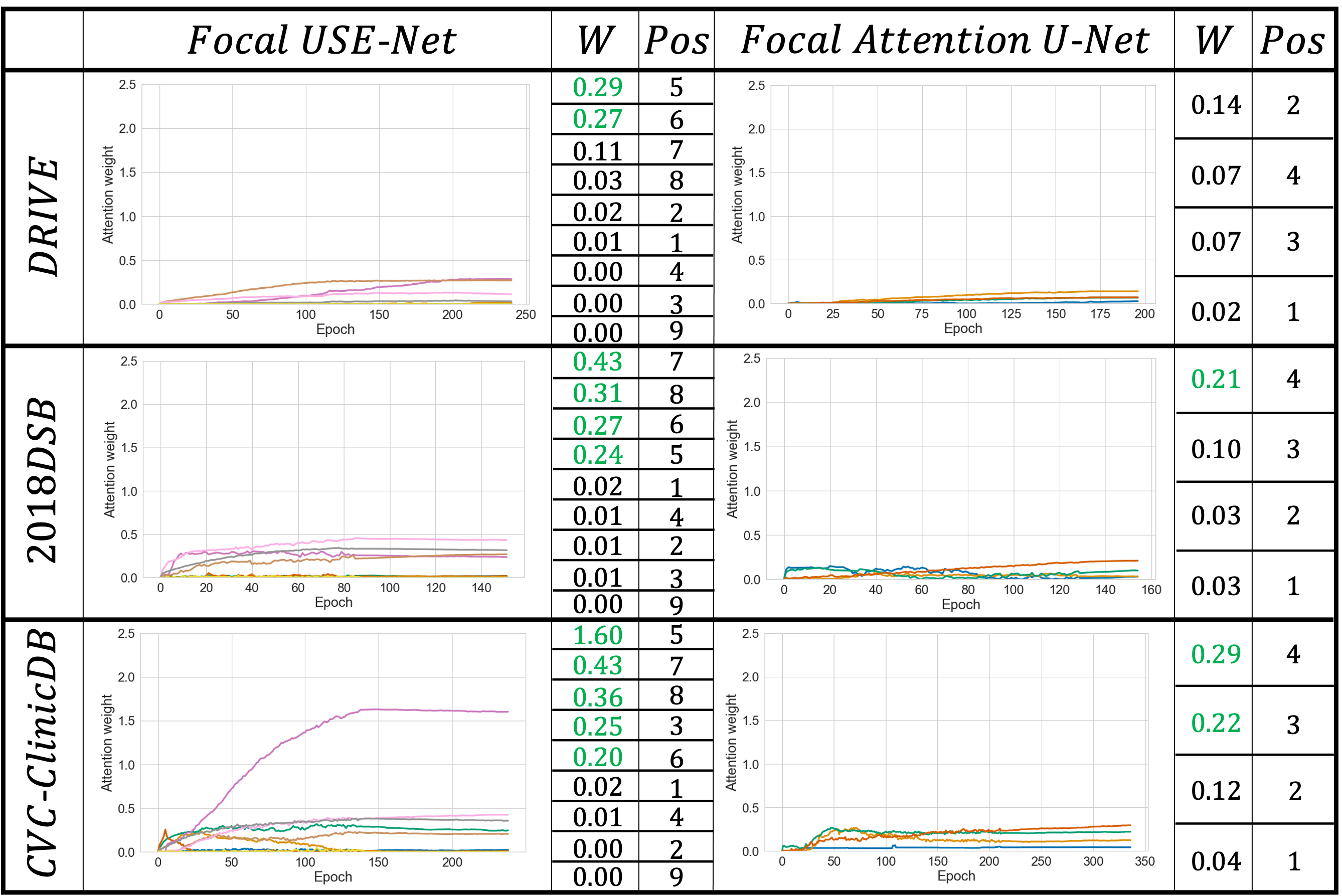}
    \caption{Zero initialised Focal layer weights monitored during training for the three datasets, with the final value, $W$, reported at convergence for each position (Pos). The colour encoding and numbering corresponds to individual attention modules illustrated in Figure 3. The attention modules with weights above $0.2$ at convergence were used in the final models and are highlighted in green.}
    \label{fig:figure_4}
\end{figure}

The Focal layer weights monitored during training, with the Focal parameter initialised to zero, are shown in Fig.~\ref{fig:figure_4}. The SE blocks in the decoder position converged towards consistently higher attention weights than in the encoder position, suggesting greater attention strength may be beneficial downstream compared to earlier in the network. The weights for both the SE and AG modules varied across datasets, with the least variation observed in the DRIVE dataset and the most in the CVC-ClinicDB dataset. This matches the variation observed within each dataset, with retinal vessels displaying considerably homogeneity in the DRIVE dataset, in comparison to colorectal polyps in the CVC-ClinicDB dataset which vary considerably in location, shape, size, colour and texture.

\section{Conclusion}

In this work, we highlighted the role of the Focal parameter in modulating attention for both loss function and network contexts. We derived a FDPT, and incorporated this into the UFL framework to generalise boundary-based losses. We demonstrated improved performance using the extended UFL over the DSC and DSC + CE losses. In the network context, we incorporated a Focal layer into the SE blocks and AG modules to optimise attention strength, and observed better performance using Focal variants of the USE-Net and Attention U-Net. Finally, we developed a simple and interpretable heuristic, by monitoring zero initialised Focal layer weights, to query the usefulness of a given attention module at a particular position. Interestingly, we often observed better performance after removing certain attention modules, suggesting that judicious use of attention modules is necessary, requiring consideration of the dataset, the model, the attention type and position, to optimise performance and efficiency. 

\clearpage

\section{Compliance with Ethical Standards}
This research study was conducted retrospectively using open-source medical imaging datasets. Ethical approval was not required as confirmed by the license attached with the open access data.

\bibliographystyle{IEEEbib}
\bibliography{strings,refs}

\end{document}